\documentstyle[amssymb,12pt]{article}
\def\be{\begin{equation}}
\def\ee{\end{equation}}
\def\L{\left}
\def\R{\right}

\input{tcilatex}

\begin{document}

\title{Gravi-electromagnetism in five dimensions and moving bodies
in Galaxy area}

 \author{W. B. Belayev\thanks{e-mail:bwb-rac@ctinet.ru} \\
   \normalsize Center for Relativity and Astrophysics,\\
   \normalsize 185 Box, 194358, Sanct-Petersburg, Russia
   \normalsize and \\ D. Yu. Tsipenyuk\thanks{e-mail:tsip@kapella.gpi.ru} \\
   \normalsize General Physics Institute of the Russian Academy of Sciences \\
   \normalsize 38 Vavilova str., Moscow, 119991, Russia}
\maketitle
\begin{abstract}
In Klein geometric model of space the mass is manifestation of the
quantized charges oscillations in additional compactified
dimension. We analyze model in which common in four-dimensional
space-time for mass and electric charge of the particle trajectory
is disintegrated in five dimensions on movement of the mass along
null geodesic line and trajectory of the charge corresponding to
the time-like interval in 5D volume. We find relation between
five-velocity vector of electric charge and mass. This scheme is
regarded to have concern with many worlds theory.

Considered approach is applied to the model of rotating space
having four-dimensional spherical symmetry. One proposed
appearance additional force in included 4D space-time, which may
be explanation of the Pioneer-effect. We analyze also possible
part of this force in conservation of the substance in Galaxy
area.
\end{abstract}

Several theories, being studied five-dimensional space-time, are
founded on different on physical principles. Generalization of the
special theory of relativity for 5D extended space
$G(T;\vec{X};S)$ with the metric (+; -, -, -, -) has been proposed
and developed in works \cite{1}-\cite{8}. Built model of extended
space (ESM) allows integration of the electromagnetic and
gravitational interactions.

The peculiarity of ESM is its studying of the particle trajectory
in 5D basing on analogy between the light movement in curved space
in general relativity and its movement in medium with refraction
coefficient being more than unity \cite{2,3,6}. In ESM mass of the
particle is component of the five-vector energy-impulse-mass in
space $G(T;\vec{X};S)$. With changes of coordinate frames in this
space the electric, gravity and scalar fields are transformed in
each other.

As the 5-th additional coordinate in ESM is used quantity, which
already exists in the (1+3)-dimensional Minkowski space
$M(T;\vec{X})$, with time coordinate $x^0=ct$, where $c$ is light
velocity, $t$ is time, and space coordinates $x^1,x^2,x^3,x^3$,
namely, interval $S$:

 \be
 \label{f1}
 S^{2}=(x^0)^{2}-(x^1)^{2}-(x^2)^{2}-(x^3)^{2}.
 \ee
This quantity is conserved at common Lorentz transformations in
the Minkowski space $M(T;\vec{X})$ but varies at turns in the
extended space $G(T;\vec{X};S)$. Thus, Minkowski space
$M(T;\vec{X})$ is a cone in extended space $G(T;\vec{X};S)$.

In 5D gravity theory, begun by Nordstr\"{o}m \cite{9} and Kaluza
\cite{10}, proposed by Klein \cite{11} approach to the particle
movement analysis examined, for example, in \cite{12}-\cite{15}
requires its trajectory to be null geodesic line in 5D space. One
ensures a particle having non-zero rest mass in the 4-D Minkowski
space to have it in 5D. The other approach to the kinematics in 5D
space is description of the movement of particle, which can be
stationary in this space and has non-zero rest mass. At that his
trajectory is determined by time-like interval
\cite{16}-\cite{20}.

In considered model integrated electromagnetism and gravity we
will turn to account following from Klein compactification
formalism founded on assumption that electric charge and mass of
observed in 4D particle have different world lines in 5D space. At
that we suppose any particle having a rest mass in 4D to contain
combination of electric charges. The mass movement conforms to
null interval, i. e.,

 \be
 \label{f2}
 0=\tilde{G}_{ij}(x_m)dx^{i}_{m}dx^{j}_{m},
 \ee
where $x^{i}_{m}$ are mass coordinates and $\tilde{G}_{ij}(x_m)$
is metric tensor of 5D space, which is function of these
coordinates. On the contrary, electric charge has trajectory with
line element

 \be
 \label{f3}
 ds^{2}_e=\tilde{G}_{ij}(x)dx^{i}dx^{j},
 \ee
where $x^{i}$ are electric charge coordinates.

Thus, world line of the particle mass in 5D intersects
corresponding world lines of electric charges and, conversely,
trajectory of the electric charge is a set of the points
appertained to trajectories of corresponding masses. This approach
leads us to proposed by Everett \cite{21}-\cite{23} conception of
many worlds founded on quantum theory.

Let us touch relation of present scheme to anthropic principle.
Our perception of environment arises by means of electro-magnetism
and we can not feel gravity direct. This may be considered as
account for assumption that to be exact mass, not electric charge
moves along null pass in 5D. In this case we can hold that 5D
space model with non-zero rest mass of the particle and
appropriate time-like interval describes electro-magnetic "trace"
of the masses, which appertain consecutively the world line of
corresponding electric charges combination. Light velocity in
extra dimension in coordinate frame of charge or system of charges
being its source is assumed to be null.

Components of mass velocity denoted as $u^{i}_m=dx^{i}_m/ds_e$
form a five-vector. With first four components of the
five-velocity vector of charge $u^i$ corresponding to interval
(\ref{f3}) equalities $u^{i}_m=u^i$ provide local coincidence of
mass and charge coordinates in 4D. Divided Eq. (\ref{f2}) on
$ds_{e}^2$ we obtain

 \be
 \label{f4}
 0=\tilde{G}_{ij}u^{i}u^{j}+(\tilde{G}_{i4}u^{i}+\tilde{G}_{4i}u^{i})u_{m}^{j}+
 \tilde{G}_{44}(u_{m}^{4})^{2}, \ i,j\neq 4.
 \ee
At this point the mass velocity along fifth coordinate will be

 \be
 \label{f5}
 u_{m}^{4}=\frac{-(\tilde{G}_{i4}+\tilde{G}_{4i})u^{i}+
 \epsilon\sqrt{(\tilde{G}_{i4}+\tilde{G}_{4i})^{2}u^{i2}-
 4\tilde{G}_{44}\tilde{G}_{ij}u^{i}u^{j}}}{2\tilde{G}_{44}}, \ i,j\neq 4,
 \ee
where $\epsilon=\pm 1$. The opposite values of $\epsilon$ conform
to matter and anti-matter. For extended Minkowski space we have

 \be
 \label{f6}
 u_{m}^{4}=\epsilon\sqrt{1+u^{42}}.
 \ee

Let us apply considered conception to analysis of metric example
of 5D space with such basis vectors that the fifth of them is not
orthogonal to others, which are basis of included 4D space. The
cosmological model with movement of the matter along fifth
coordinate based oneself on metric conforming to this property has
been studied in \cite{20}. We analyze space-time, including
four-dimensional spherical space, with coordinates
$x^i=(ct,a,\theta,\varphi,\chi)$ to be rendered to orthogonal
frame by transformation
\begin{eqnarray}
\label{f7}
 \eta_0=ct, \nonumber \\
 \eta_1=a\cdot\sin\chi\cdot\sin\theta\cdot\cos\varphi, \nonumber \\
 \eta_2=a\cdot\sin\chi\cdot\sin\theta\cdot\sin\varphi,\\
 \eta_3=a\cdot\sin\chi\cdot\cos\theta, \nonumber \\
 \eta_4=a\cdot\cos\chi. \nonumber
\end{eqnarray}
This space is assumed to be rotating and the metric is taken in
form
 \be
 \label{f8}
 ds^2=c^2[1-a^2B(a)^2]dt^2-da^2-a^2[2cB(a)dtd\chi+\sin^2\chi(d\theta^2+
 \sin^2\theta d\varphi^2)+d\chi^2],
 \ee
where $B(a)$ is dependent on $a$ coefficient. In accordance with
considered approach with $ds>0$ conforms to the movement of
particle's electro-magnetic trace.

In \cite{24} the geodesic line equations was written in form

 \be
 \label{f9}
 \frac{d}{ds}(\tilde{G}_{ij}u^{j})-\frac{1}{2}\frac{
 \partial\tilde{G}_{mj}}{\partial x_{i}}u^{m}u^{j}=0.
 \ee
For metric (\ref{f8}) zeroth, first and fourth components of these
equations with comoving coordinates of convenional type
($u^{1}=u^{2}=u^{3}=0$) yield
\begin{eqnarray}
\label{f10}
 \frac{d}{ds}[(1-a^2B^2)u^0-a^2Bu^4]=0,  \\
 \label{f11}
 aB\L( B+a\frac{\partial B}{\partial a}\R)u^{02}+a\L(2B+a\frac{\partial B}
 {\partial a}\R)u^0u^4+au^{42}=0,  \\
 \label{f12}
 \frac{d}{ds}[a^2Bu^0+a^2u^4]=0.
\end{eqnarray}
Solution of this system must be compatible with set by metric
(\ref{f8}) condition

 \be
 \label{f13}
 u^{02}-a^2(u^{4}+Bu^{0})^2=1.
 \ee
Such solution will be

 \be
 \label{f14}
 u^{0}=\xi, \ u^{4}=-\xi B(a),
 \ee
where $\xi$ takes values $1$ and $-1$. Corresponding fifth
component of mass five-velocity (\ref{f5}) is following:

 \be
 \label{f15}
 u_{m}^{4}=\xi\L[\frac{\epsilon}{a}+B(a)\R].
 \ee
We notice that this equation with other four components of
$u_{m}^i$ does not put null geodesics of particle mass but it is
only velocity of mass in every point of charge geodesics.

Considered coordinate frame is transformed to coordinates
$x^i=(ct,r,\theta,\varphi,y)$ having only 3D symmetry by
expressions
\begin{eqnarray}
\label{f16}
 r=a\cdot\cos\chi, \nonumber \\
 y=a\cdot\sin\chi.
\end{eqnarray}
We find acceleration $d^{2}r/ds^2$ with condition of particle
geodesic movement set by Eqs. (\ref{f14}) when the fifth
coordinate is $\chi=0$. In this case we obtain $y=0$ and
$dr/ds=0$. The fifth component of the five-velocity is written as

 \be
 \label{f17}
 u^{4}=\frac{1}{r^2+y^2}\L(r\frac{dy}{ds}-y\frac{dr}{ds}\R).
 \ee
This equation yields

 \be
 \label{f18}
 \frac{dy}{ds}=-\xi\cdot rB(r).
 \ee
Equation of the system (\ref{f9}) with $i=1$ is rewritten as

 \be
 \label{f18}
 \frac{d^{2}a}{ds^2}=0.
 \ee
After some transformations we obtain

 \be
 \label{f20}
 \frac{d^{2}r}{ds^2}=-rB(r)^2.
 \ee
Thus, rotation in 5D gives additional force in included 4D space.
This force is invariable with

 \be
 \label{f21}
 B(a)=Ka^{-1/2},
 \ee
where $K$ is constant. One may be considered as explanation of
additional acceleration of the Pioneer 10 $a_p=8.5\cdot10^{-8} \rm
{cm/s^2}$ towards to the receiving antenna on the Earth
\cite{25}-\cite{27}. Assumed interval to be $ds=cdt$ we obtain
value of the constant which is $K=0.97\cdot10^{-13} \rm
{cm^{-1/2}}$.

This additional acceleration satisfies to the constraint
$a_{p}R\gtrsim v_{sw}^2$, where $R$ is Galaxy radius and $v_{sw}$
is velocity of the solar wind \cite{28}. It follows from this that
if Pioneer-effect spreads on whole Galaxy area, it is conducive to
conservation of the matter within the bounds of Galaxy belt.


\small
\begin{thebibliography} {28}

\bibitem{1} D.Yu. Tsipenyuk, V.A. Andreev, {\it Krattkie Soobstcheniya po
Fizike\/} (Bulletin of Lebedev Physics Institute (Russian Academy
of Sciences)) {\bf 6}, 23--34, (2000); arXiv: gr-qc/0106093.

\bibitem{2} D.Yu. Tsipenyuk, V.A. Andreev, {\it Issledovano v Rossii\/},
{\bf 60}, (1999) (in Russian);
http://zhurnal.ape.relarn.ru/articles/1999/060.pdf .

\bibitem{3} D.Yu. Tsipenyuk, {\it Gravitation and
Cosmology\/}, Vol. {\bf 7}, No.{\bf 4}(28), 336--338, (2001);
arXiv: physics/0203017.

\bibitem{4} D.Yu. Tsipenyuk, {\it Krattkie Soobstcheniya po Fizike\/}
(Bulletin of Lebedev Physics Institute (Russian Academy of
Sciences)) {\bf 7}, 39--49, (2001); arXiv: physics/0107007.

\bibitem{5} D.Yu. Tsipenyuk, V.A. Andreev, {\it Krattkie Soobstcheniya po
Fizike\/}(Bulletin of Lebedev Physics Institute (Russian Academy
of Sciences)) {\bf 6}, 3--15, (2002); arXiv: physics/0302006.

\bibitem{6} D.Yu. Tsipenyuk, {\it Issledovano v Rossii\/} {\bf 81}, 907--916,
(2001)(in Russian);
http://zhurnal.ape.relarn.ru/articles/2001/081.pdf .

\bibitem{7} D.Yu. Tsipenyuk, V.A. Andreev,  "Electrodynamics in Extended
Space", {\it preprint IOFAN\/} {\bf 9}, Moscow, (1999)(in
Russian).

\bibitem{8} D.Yu. Tsipenyuk, V.A. Andreev, "Gravitational effects in
Extended Space", {\it preprint IOFAN\/} {\bf 4}, Moscow,(2001)(in
Russian).

\bibitem{9} G. Nordstrom, {\it Phyz. Zeitschr.\/}{\bf 1}, 504 (1914)
(in German).

\bibitem{10} T. Kaluza, {\it Sitz. Preuss. Akad. Wiss.\/}
 Phys. Math. K{\bf 1}, 966 (1921) (in German).

\bibitem{11} O. Klein {\it Z. Phys.\/} {\bf 37}, 895 (1926).

\bibitem{12} J. Van Dongen, arXiv: gr-qc/0009087.

\bibitem{13} P.S. Wesson, S.S. ~Seahra, {\it Gen. Rel. Grav.\/}, {\bf 33},
1731 (2001); arXiv: gr-qc/0105041.

\bibitem{14} W.B. Belayev, "Extra force in Kaluza-Klein gravity theory",
Invited talk at "Gravitation, Cosmology and Relativistic
Astrophysics" (Kharkiv, Ukraine, June 23-27, 2003);
 arXiv: gr-qc/0308076.

\bibitem{15} P.S. Wesson, {\it Gen. Rel. Grav.\/} {\bf 35}, 307
(2001); arXiv: gr-qc/0302092.

\bibitem{16} J.M. Overduin, P.S. Wesson, {\it Phus. Rept.\/}, {\bf 283}, 303
(1997); arXiv: gr-qc/9805018.

\bibitem{17} J. Ponce de Leon, {\it Phys. Lett. B\/} {\bf 523}, 311 (2001);
arXiv: gr-qc/0110063.

\bibitem{18} R. Maartens, "Geometry and dynamics of the brane-world",
invited talk at EREs2000, Spanish Relativity Meeting; arXiv:
gr-qc/0101059.

\bibitem{19} F. Dahia, E.M. ~Monte, C. ~Romero, {\it Mod. Phys. Lett. A\/}
{\bf 25}, 1173, (2003); arXiv: gr-qc/0303044.

\bibitem{20} W.B. Belayev, {\it Spactime and Substance\/} {\bf 7}, 63 (2001);
arXiv: gr-qc/0110099.

\bibitem{21} B.S. DeWitt, R. N. ~Graham, "The many-worlds
Interpretation of Quantum Mechanics", eds. R. N. ~Graham,
Princeton Series in Physics, Princeton University Press (1973).

\bibitem{22} F.J. Tipler, in "Quantum Concepts of Space and
Time", eds R. Penrose and C. Isham, Oxford University Press
(1986).

\bibitem{23} D. Bohm, B.J. Hiley, "The Undivided Universe", Routledge
Press, London, 1993.

\bibitem{24} G.C. McVittie, "General Relativity and Cosmology"
Chapman and Half Ltd., London, 1956. (Russian translation: G.C.
Mac-Vitti, "General Relativity and Cosmology", IIL, Moskow, 1961.)

\bibitem{25} J.D. Anderson, P.A. Laing, E.L. Lau, A.S. Liu, M.M. Nieto,
S.G. Turushev, {\it Phys. Rev. Lett.\/} {\bf 81}, 2858 (1998);
arXiv: gr-qc/9808081.

\bibitem{26} S.G. Turushev, J.D. Anderson, P.A. Laing, E.L. Lau,
A.S. Liu, M.M. Nieto, "The apparent anomalous, weak, long-range
acceleration of Pioneer 10 and 11", XXXIV-th Rencontres de Morion
Meeting on Gravitational Waves and Experimental Gravity 1999, Les
Arcs, Savoi, France; arXiv: gr-qc/9903024.

\bibitem{27} J.D. Anderson, P.A. Laing, E.L. Lau, A.S. Liu, M.M. Nieto,
S.G. Turushev, {\it Phys. Rev. D\/} {\bf 65}, 082004 (2002);
arXiv: gr-qc/0104064.

\bibitem{28} K. Tren\v{c}evski, unpublished data, 2004.

\end{thebibliography}
\end{document}